\documentclass[useAMS,usenatbib,usegraphicx]{mn2e}

\def\asca{{\sl ASCA}}
\def\sax{{\sl BeppoSAX}}
\def\xmm{{\sl XMM-Newton}}
\def\c{{\sl Chandra}}
\def\rosat{{\sl ROSAT}}
\def\sirtf{{\sl SIRTF}}
\def\iso{{\sl ISO}}
\def\iras{{\sl IRAS}}
\def\ltsim{\mathrel{\hbox{\rlap{\hbox{\lower4pt\hbox{$\sim$}}}\hbox{$<$}}}}
\def\gtsim{\mathrel{\hbox{\rlap{\hbox{\lower4pt\hbox{$\sim$}}}\hbox{$>$}}}}
\def\ergpspsqcm{ erg s$^{-1}$ cm$^{-2}$}
\def\ergps{ erg s$^{-1}$}

\def\Msun{M$_{\odot}$}
\def\micron{$\mu$m}

\def\lognh{log$N_{\rm H}$}
\def\nh{$N_{\rm H}$}
\def\zmax{$z_{\rm max}$}
\def\zcut{$z_{\rm cut}$}
\def\pl{$p_L$}
\def\pn{$p_n$}

\title{X-ray background synthesis: the infrared connection}

\author[P. Gandhi and A.C. Fabian]
{\parbox[]{6.in} { P. Gandhi$^{\dag}$ and A.C. Fabian\\
\footnotesize
Institute of Astronomy, Madingley Road, Cambridge CB3 0HA, UK \\
$^{\dag}$pgandhi@eso.org\\}}
\date{Accepted 6 Nov 2002.
      Received 9 Aug 2002}
\pubyear{}

\begin{document}

\maketitle

\begin{abstract}
We present a synthesis model of the X-ray background based on the cross-correlation between mid-infrared and X-ray surveys, where the distribution of type 2 sources is assumed to follow that of luminous infrared galaxies while type 1 sources are traced by the observed \rosat\ distribution. The best fits to both the X-ray number counts and background spectrum require at least some density evolution. We explore a limited range of parameter space for the evolutionary variables of the type 2 luminosity function. Matching the redshift distribution to that observed in deep \c\ and \xmm\ fields, we find weak residuals as a signature of Fe emission from sources in a relatively peaked range of redshift. This extends the recent work of \citeauthor{franceschini02}, and emphasizes the possible correlation between obscured AGN and star-forming activity.

\end{abstract}

\section{Introduction}

Intensive follow-up work of recent deep \c\ and \xmm\ fields has shown that, after 40 years, the X-ray background (XRB) still holds surprises in store. While the bulk of the 2--10 keV background has been resolved into discrete sources (mostly active galactic nuclei; AGN), the majority of the power contribution to the XRB may emerge at relatively low redshifts of $z<1$ \citep{rosati02, hasinger02, brandt02}. This is in contrast to the distribution of unobscured X-ray selected AGN which peaks at $z\ge 1.5$ (e.g., \citealt{boyle93}; \citealt{mi00}), the evolution of QSOs observed in the optical and its decline only at high redshifts \citep{boyle00, fan01} and XRB synthesis models which have, until now, predicted a similar peak at higher $z$ (while successfully reproducing the XRB spectrum; e.g., \citealt{settiwoltjer89}, \citealt{comastri95}). Of course, underlying these connections is the orientation-dependent unified model of AGN.

Increasingly sophisticated models have been able to increase their goodness of fit by forcing fast evolution for absorbed sources to lower redshift (e.g., \citealt{gilli01} use a fast evolution to $z$=1.3). However, it is now likely that the characteristic source redshift is as low as $z$$\sim$0.7 (\citealt{hasinger02}; even after accounting for the effects of large scale structure).



On the other hand, the luminosity function (LF) of infrared galaxies observed by missions such as the InfraRed Astronomy Satellite (\iras) and the Infrared Space Observatory (\iso) has been recently inferred to evolve steeply to $z\sim 0.8$ and flatten thereafter (e.g., \citealt{ce01}; \citealt{franceschini01_long}). While this regime is dominated by obscured star-formation, reprocessed AGN accretion activity may also have a non-negligible contribution. \citet{fadda01} found that (while the fraction of mid-infrared sources with X-ray counterparts is more than 10 per cent) at least 30 per cent of X-ray sources in the Lockman Hole and Hubble Deep Field-North have mid-infrared (MIR) \iso\ counterparts, and this fraction increases to 63 per cent for the sources detected in the 5--10 keV band. Radiative transfer modelling of the broad-band spectral energy distribution (SED) of a number of \iso\ sources has suggested that absorbed AGN emission can indeed account for observed far-infrared SED peaks \citep{wfg, franceschini02b}. This evidence suggests that X-ray \lq type 2\rq\ Seyferts and quasars may be ubiquitously detected in the 10--100 \micron\ regime. 

Here we present a synthesis model for the XRB and X-ray number counts, based on the assumption that type 2 obscured AGN are traced by a LF which follows the infrared distribution. Type 1 unobscured X-ray sources, on the other hand, are distributed according to the well-determined soft X-ray luminosity function (XLF). We include a distribution of source spectra, accounting for full Thomson scattering cross-sections and explore a range of parameter space for volume emissivity evolution with redshift. We comment on the need for various classes of sources in matching the observational constraints, and investigate the effects of including an Fe K$\alpha$ emission line in the template spectra. This follows on from the work of \citet[][ hereafter F02]{franceschini02}, who adopted a fixed fraction of a composite MIR luminosity function for the density and evolution of a single type 2 X-ray SED, and were successful at reproducing the observed peak of the redshift distribution in deep fields. Finally, we construct far-infrared SEDs under the assumption of complete reprocessing of the AGN blue bump radiation to predict obscured AGN source counts at 70 \micron, and discuss implications for unification models.


For the purposes of this work, we ignore distinctions in the optical and X-ray classifications of AGN: i.e., type 1 $\equiv$ unobscured (or mildly obscured, see next section) and type 2 $\equiv$ obscured. We assume H$_0$=50 km s$^{-1}$ Mpc$^{-1}$, $\Omega_{\rm M}$=1 and $\Omega_{\Lambda}$=0.

\section{The model}
\subsection{Spectra}
\label{sec:spectra}
The AGN X-ray spectra are assumed to be power-laws with a slope fixed at $\alpha$=0.9 ($F_E\propto E^{-\alpha}$) and an exponential cutoff at $E_C=360$ keV. A reflection hump from cold material at solar abundance is added by employing the PEXRAV code \citep{pexrav}, assuming uniform 2$\pi$ reflection and inclination angle $\theta=60^{\circ}$. Iron emission lines are not included here, and are discussed separately (\S~\ref{sec:FeK}).

Obscuration by material of solar abundance local to the source is modelled (with \lognh=20...25; step 0.5) using the cross-sections of \citet{mm83}. The importance of Compton down-scattering for Thomson-depths ($\tau$) $\gtsim$1 (\lognh$\gtsim$24) has been discussed previously [\citet[][ hereafter C95]{celotti95}; \citet[][ hereafter WF99]{wf}] and we model the full Klein-Nishina cross-section using the same monte carlo code as the latter authors, again assuming solar abundance.

Finally, 2 per cent of the intrinsic continuum is added to all spectra to represent the component scattered into the line-of-sight. See Fig~1 of WF99 for an illustration of the constructed spectra. 

The ratio of type 2 : type 1 sources in the local Universe is thought to be around 4 and could be larger than 10 \citep{maiolinorieke95,matt00}. We parametrize this distribution as (\lognh)$^\beta$. Type 1 sources are taken to be those with \lognh$\le$22, with the rest being type 2. We note that this renders our model effectively \lq unified\rq, since such a parametrization can be expected from the angular distribution of torus column density, where the maximum optical depth is along the equatorial plane with the most probable orientation (discussed in C95). However, this qualitatively matches the increasing fraction of highly obscured sources (at least to \lognh=24) observed by \citet{risaliti99} and also allows us to explore a range of possible distributions.



\subsection{Luminosity functions and their evolution}
\label{sec:llf}

The type 1 XLF that we use is the observed-frame 0.5--2 keV double power-law form found by \citet{mi00}, with a density evolution which is luminosity-dependent. The parameters of the LF (both local and evolutionary) that we adopt are those of the BAS1 model quoted in \citet{mi00b}. The only change is the relative weighting of type 1 sub-classes (with different $N_{\rm H}$) as described above.

The LF of type 2 X-ray sources is completely unknown, due to the difficulty of compiling an unbiased and complete sample.
As mentioned, we assume that these are traced by a MIR local luminosity function (LLF). A number of such LLFs have been determined, and here we adopt the one found at 15 \micron\ by \citet{xu98}, based on incorporation of \iso-determined SEDs into 1406 \iras\ galaxy observations. \citet{ce01} and \citet{franceschini01_long} have used this LLF (partially or fully, in combination with other LLFs) to successfully reproduce the slope of the observed MIR number counts at faint fluxes with appropriate evolution. We have not explored the dependence on the various LLFs, since our purpose here is only to demonstrate the viability of the X-ray:IR connection.


There are then two major unknowns in this work, for which there are no firm observational constraints: 1) the relative X-ray:IR intrinsic luminosity ratio for each SED; 2) the fraction of IR sources which are X-ray detected AGN, as a function of \lognh. The observed X-ray:IR ratio will depend on the amount of obscuration and larger absorbing columns of gas and dust will deplete more X-rays and reprocess larger amounts of optical/UV radiation to the IR. Using the standard \citet{elvis94} median SED for radio-quiet quasars (with the normalization of the LF discussed below) leads to an underestimation of the XRB spectrum by as much as a factor of 3 (or overestimation of the counts, if we correspondingly correct the LF upwards). However, there are suggestions that the X-ray:IR ratio can be similarly large in the case of Seyferts. This can also be interpreted as assuming that the bolometric correction factor (i.e., $L_{2-10}/L_{\rm Bol}$) is closer to 10 per cent, as compared to 3 per cent for quasars, where $L_{2-10}$ is the {\sl intrinsic} X-ray luminosity and $L_{\rm Bol}$ is the bolometric luminosity. For instance, the general trend of luminosity ratios for the sample of \citet[][ see their Fig~9]{sanders89} suggests this possibility. Indeed, if the XRB is dominated by Seyferts as opposed to quasars, then a larger bolometric correction may be appropriate. With this approximation, we find that $L_{2-10}\approx L_{\rm 15\mu m}$. 

\begin{figure}
  \begin{center}
\includegraphics[angle=90,width=8.5cm]{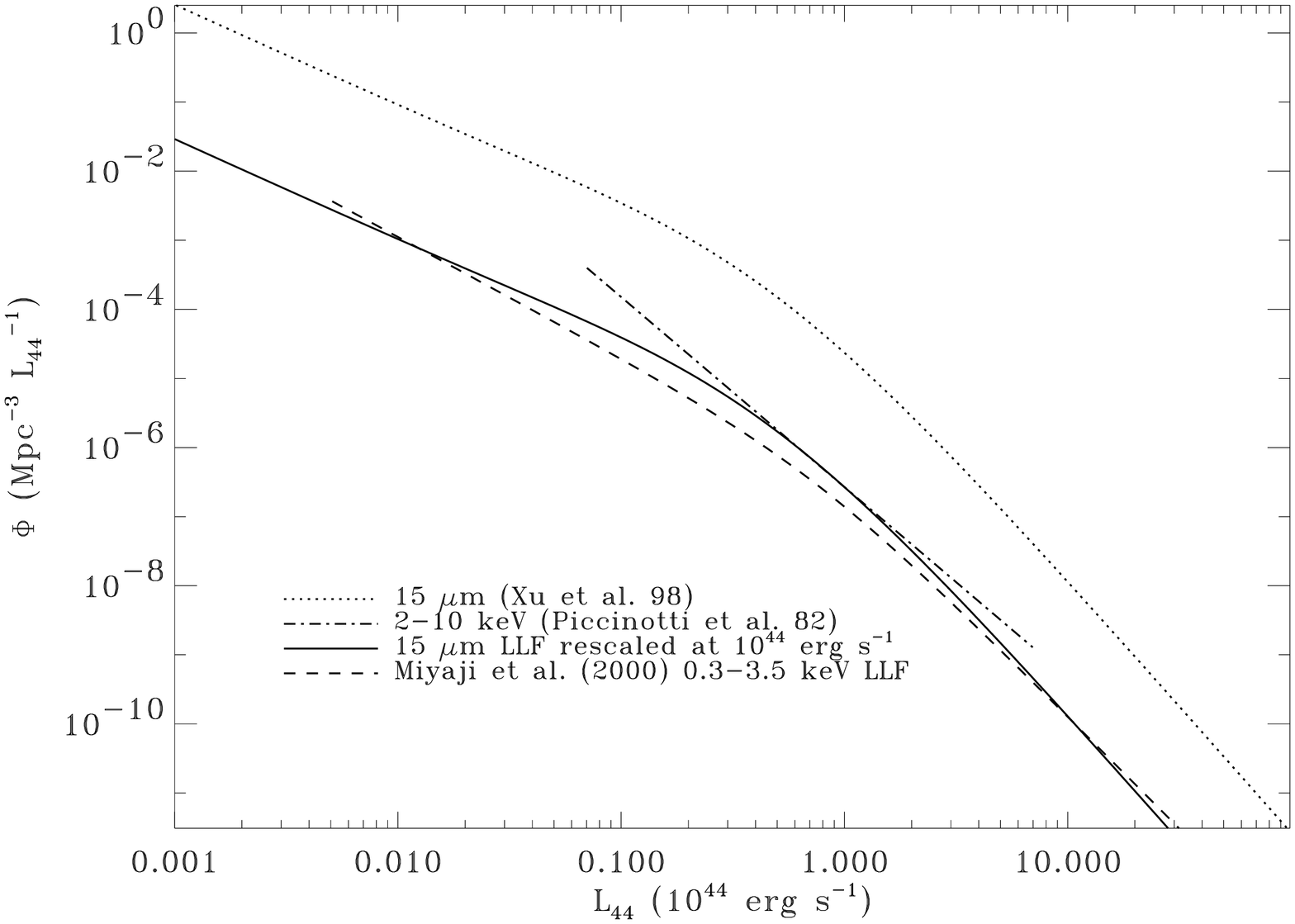}
  \caption{The local luminosity functions relevant to this paper. The dotted line is the 15 \micron\ LLF of \citet{xu98}. This is normalised to the 2--10 keV X-ray LLF of \citet[][ dot-dashed line]{piccinotti82} at 10$^{44}$ \ergps\ -- the result is the solid line, which is the adopted type 2 LLF for each \nh\ sub-class {\sl without} the appropriate weighting factors. The combined LLF for all the type 2 sources will depend on the type 2 : type 1 ratio, which will vary with $\beta$. The dashed line is the adopted LLF of type 1 sources in the observed 0.5--2 keV band from \citet{mi00}.} \label{fig:llfs} \end{center}
\end{figure}

To determine the fraction of IR sources which will be X-ray detected AGN (the rest are probably obscured starbursts, as discussed in the introduction), we choose to match the MIR LF to the X-ray LF determined in the 2--10 keV band by \citet{piccinotti82}, and adopt this as the XLF for each $N_{\rm H}$ sub-class. We do this by simply scaling so that the two match at 10$^{44}$ \ergps. No SED corrections are needed, since the 2-10 keV and 15\micron\ luminosities are approximately equal. The adopted LLFs are shown in Fig~\ref{fig:llfs}. While this prescription may be rather crude, it is self-consistent in the sense that the local X-ray space density (from all $N_{\rm H}$ sub-classes) adds to a few per cent of the IR space-density, as found by \citet[][ discussed in the following sections]{fadda01}.

For the evolution with redshift of the type 2 LLF, we investigate power-law emissivity evolution [as $(1+z)^{p}$] out to \zcut\ and a constant weighting thereafter to \zmax\ (set equal to the weight at \zcut).


\subsection{Model fitting}

Integration has been carried out from log$L_{\rm min}$=41.7 to log$L_{\rm max}$=47.5 (the limits of \citealt{mi00}; $L$ refers to the observed 0.5--2 keV band for type 1 sources and the rest-frame 2--10 keV band for type 2 sources in units of erg s$^{-1}$) for all the source sub-classes, with the variables being the type 2 parameters: \zmax, \zcut\ and $p$. In accordance with C95 and WF99, we normalize the generated spectrum by a factor (determined during fitting) between 0.7 and 1.5, in order to account for uncertainties in the absolute level of the background, deviations from the \lq true\rq\ luminosity function and degeneracy of cosmology. A good fit is assumed if the deviation of the normalized spectrum on a grid of energies at 10, 20, 40, 50, 75 and 100 keV is $\ltsim$10 per cent, while a maximum contribution of 97 per cent at 5 keV is required, in order to allow for contributions due to clusters, which we do not explicitly model here.

We further discuss a model that provides a good description of the X-ray number counts and redshift distribution. However, an explicit fit is not performed due to the number of parameters involved. 

\section{Results}

\subsection{Matching observational constraints}

\label{section:results1}

\begin{figure*}
  \begin{center}
\includegraphics[angle=90,width=12cm]{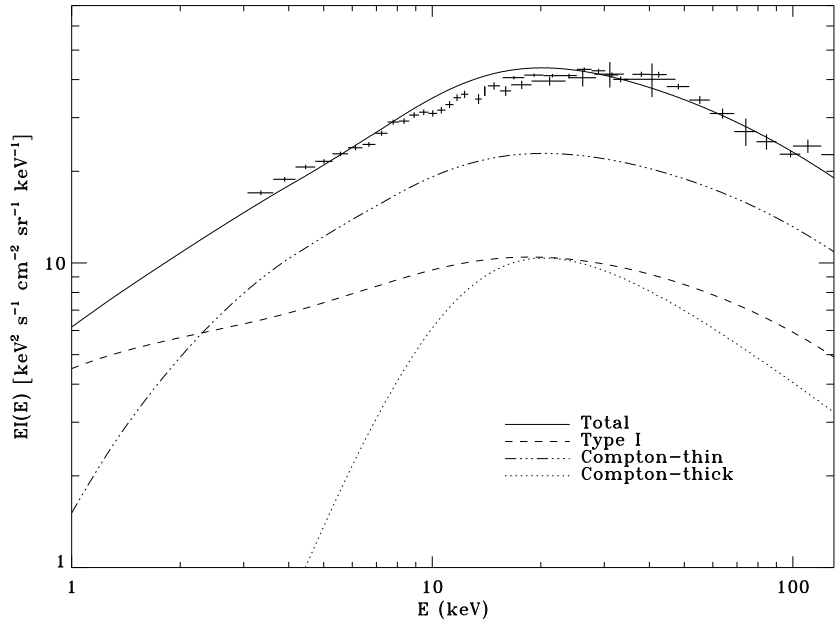}
  \caption{The crosses represent the observed XRB spectrum from \citet{gruber99}. Our model has been fit to the analytic form from the same paper between $\sim$10 and 100 keV. The solid line is the total spectrum with a normalization factor of 1.44. Contributions from type 1 sources (dashed), type 2 Compton-thin sources with 22$\le$\lognh$<$24 (dot-dot-dot-dashed) and Compton-thick sources (dotted) are shown.} \label{fig:xrbspec} \end{center}
\end{figure*}

The fit to the XRB spectrum for one of our \lq good\rq\ models is shown in Fig~\ref{fig:xrbspec}, for which \zcut=0.7, \zmax=4.5 and $p$=4, with a normalization factor of 1.44. 
The contribution from various classes of AGN is shown as well, with type 1 sources dominating below 2 keV, while Compton-thin type 2 sources dominate beyond. Compton-thick sources with \lognh$>$24 never dominate the XRB due to the effect of Thomson down-scattering.

\begin{figure}
  \begin{center}
\includegraphics[angle=90,width=8.5cm]{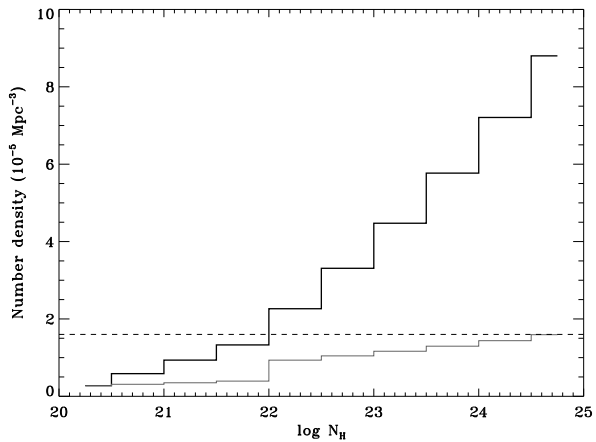}
  \caption{Local space density of sources implied by the model in Fig~\ref{fig:xrbspec}. The light, thin line shows the differential density per unit \lognh. The solid line is the cumulative density. For comparison, the number density predicted by the \citet{piccinotti82} XLF (integrated over the luminosity limits stated therein) is shown as the dashed line.} \label{fig:ndensity} \end{center}
\end{figure}

The local number density as a function of \lognh\ required by this model is shown in Fig~\ref{fig:ndensity}. The weighting parameter ($\beta$) = 5, implying a local type 2 : type 1 ratio of 5.6. The local 2--10 keV volume emissivity implied is 4.5$\times$10$^{38}$\ergps\ Mpc$^{-3}$, in agreement with the values derived by \citet{mi94} by cross-correlating the XRB with \iras\ galaxies. 

\begin{figure}
  \begin{center}
\includegraphics[angle=90,width=8.5cm]{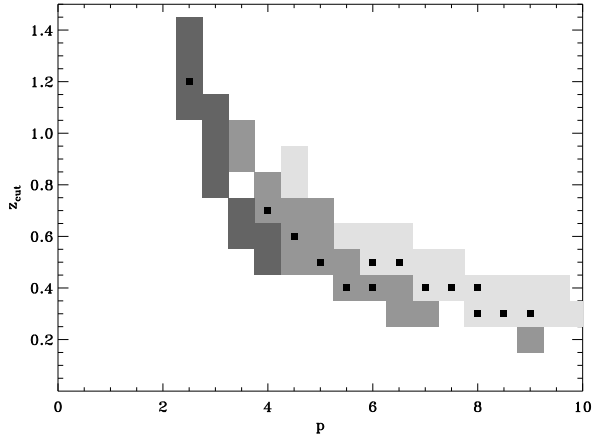}
  \caption{The range of acceptable models which fit the XRB spectrum only. The shades of grey represent different values of $N_{\rm H}$ weighting index $\beta$. Lightest: $\beta$=2; Intermediate: $\beta$=5; Darkest: $\beta$=8. These represent deviations from the observed spectrum of less than 12 per cent at 10, 20, 40, 50, 75 and 100 keV, after normalization as described in the text. The filled black squares represent models with deviations of less than 10 per cent. \zmax\ values for these models typically range from 3--5.} \label{fig:grid} \end{center}
\end{figure}

More generally, the region of parameter space explored is shown in Fig~\ref{fig:grid}. For the models shown, \zmax\ values of 0.5...5 were explored. The shaded regions show models which deviate from the analytic fit to the observed XRB determined by \citet{gruber99} by at most 12 per cent at the energy values mentioned.
The smaller black squares locate models where the maximum deviation is 10 per cent. The range of acceptable models is relatively narrow, since advancing along either axis corresponds to increasing source weighting. 
WF99 pointed out that parameter space could be opened up by shifting the individual spectral peaks above 30 keV by allowing super-solar metallicities. Including the effects of coronal Comptonization of the reflection hump may also help to a small extent \citep{petrucci01}.

The other observational constraint that we match is the integral number counts in the 2--10 keV band. We have explored splitting emissivity evolution into luminosity and density evolution (with power-law indices \pl\ and \pn\ respectively) without affecting the XRB spectrum itself. While there is degeneracy between these possibilities, we find that some density evolution is required to delay the flattening of the integral number counts to below 10$^{-16}$ \ergpspsqcm\ (as also pointed out by F02). Models with pure luminosity evolution produce a much flatter slope in the counts at faint fluxes. We present one model with \pl=2 and \pn=2 in Fig~\ref{fig:counts210}. The observed counts are from \citet{rosati02}.

The redshift distribution of sources predicted by this model is shown in Fig~\ref{fig:nz}, where the histogram is the (unnormalised) source distribution from \citet[][ their Fig 6]{hasinger02}. The spectroscopic completeness fraction reported by Hasinger is about 60 per cent -- the rest are likely to be type 2 sources at intermediate to high redshifts with weak (or absent) emission lines. Low redshift sources are likely to have a high identification completeness. We have thus normalized the observed distribution to the model distribution at $z=0.5$ (avoiding $z=0.7$ due to the presence of large scale structure). The fraction of identified sources beyond $z=1$ is about 40 per cent of the total predicted number. Exact comparison is difficult due to the mixture of flux limits of the surveys from which the identified sources are drawn.
As also mentioned by F02, at least part of the shortfall of sources at $z\ltsim 0.5$ could be due to the contribution by clusters.

\subsection{The need for highly obscured sources}

\begin{figure*}
  \begin{center}
\includegraphics[angle=90,width=12cm]{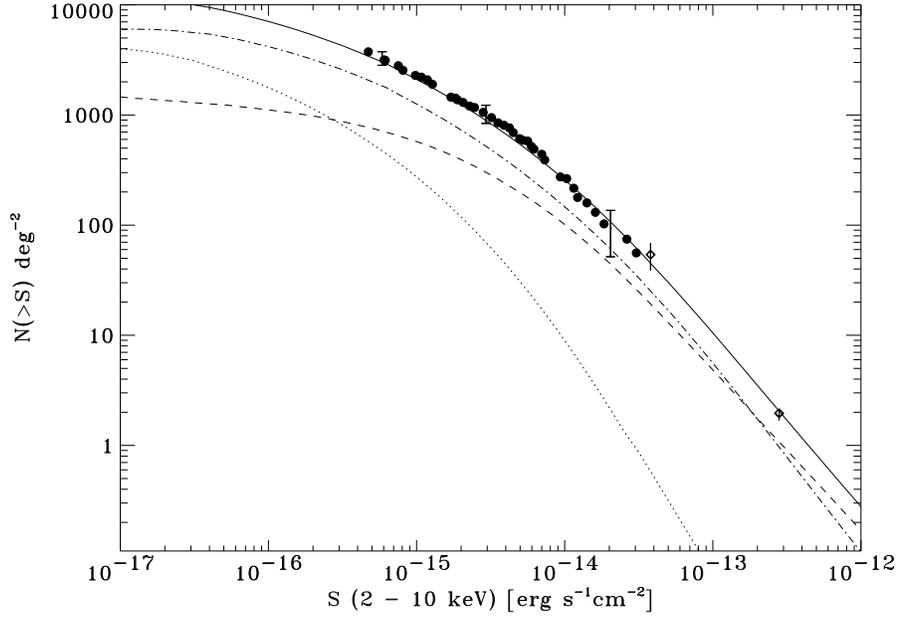}
  \caption{2--10 keV source counts predicted by the model of Fig~\ref{fig:xrbspec} (solid line), with contributions from various source types similarly shown. The solid circles are a sample of the source counts reported by \citet{rosati02}. Three representative error bars (with hats) are shown. The diamonds with error-bars are from (with increasing flux) \asca\ \citep{ogasaka98} and \sax\ \citep{giommi00}.} \label{fig:counts210} \end{center}
\end{figure*}

\begin{figure*}
  \begin{center}
\includegraphics[angle=90,width=12cm]{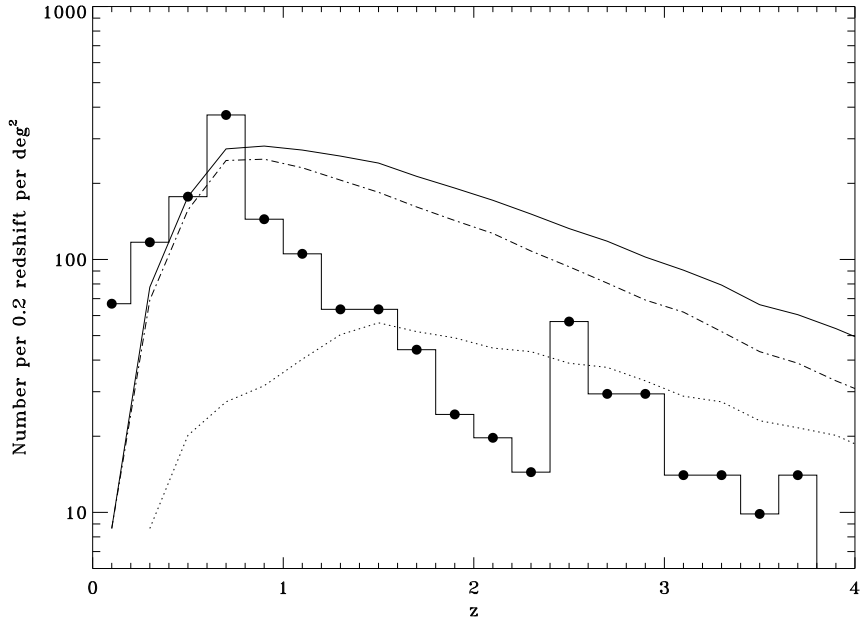}
  \caption{Source distribution as a function of redshift, calculated in steps of 0.2 in z (solid line) for a flux limit S(2--10 keV)$>$5$\times$10$^{-16}$ \ergpspsqcm. The contribution of type 1 sources is shown as the dotted line, while that of all type 2 sources is shown as the dot-dashed line. The observed distribution of \citet[][ histogram and black circles]{hasinger02} has been normalised to the predicted number at $z=0.5$ (see text).} \label{fig:nz} \end{center}
\end{figure*}

The major discrepancy between the observed and the modelled XRB spectrum occurs between 10 and 20 keV, close to the position of the redshifted peak (from 30 keV) of a source at the characteristic redshift of 0.7 (Fig~\ref{fig:xrbspec}). The maximum deviation at 14 keV is slightly more than 10 per cent. Although there is no doubt of the need for obscured accretion in producing the hard spectrum of the XRB, the relatively sharp peak of the Compton-thick contribution (\lognh$>$24) at $\sim$20 keV (dotted line) suggests that removing these sources may decrease the maximum percentage deviation of the model. We have tried setting log$N_{\rm H}^{\rm max}$=24 and, indeed, find many solutions which fit the XRB spectrum well. However, removal of these sources effectively makes the total spectrum softer, and thus forces a much lower \zmax\ (with appropriate renormalization) to reproduce the flat slope below $\sim$30 keV. This leads to number counts being severely underestimated. In addition, the contribution of Compton-thick sources is important to the 2--10 keV source counts at the very faintest fluxes (Fig~\ref{fig:counts210}), and this should be even more so for counts at harder energies. Thus, they cannot be easily excluded from the model.

We also note that marginally acceptable models with no type 2 quasars (i.e. sources with \lognh$>$22 and log$L$$>$44) can be found. Their contribution is relatively small (due to the decrease in their number density by $\gtsim$$10^4$ compared to log$L$=42; Fig~\ref{fig:llfs}), mainly showing up in the number counts at brighter fluxes above $10^{-15}$ \ergpspsqcm.

\subsection{Features in the spectrum}
\label{sec:FeK}

One consequence of having a \lq characteristic\rq\ source redshift around which the bulk of the background is produced is that an imprint may be left on the background spectrum itself. No such characteristic features (including any Fe emission lines) have been found to date (e.g., \citealt{schwartz92, fabianbarcons92} and references therein) and this has been taken as strong evidence for a truly cosmological population of sources which is well spread-out in redshift, thereby smearing out any features. We investigate the effect of incorporating an Fe K$\alpha$ emission line with our relatively-peaked redshift distribution.



\begin{figure}
  \begin{center}
\includegraphics[angle=90,width=8.5cm]{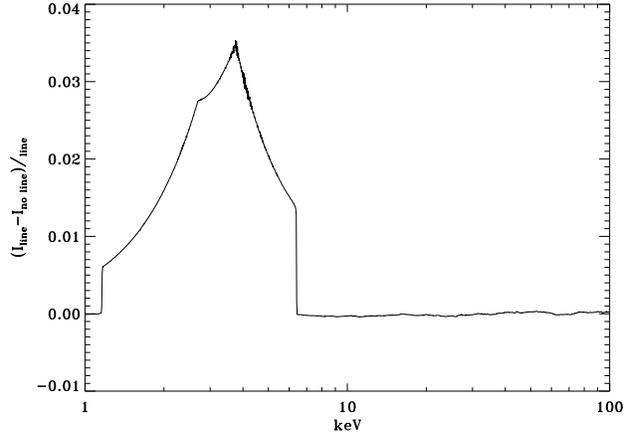}
  \caption{
Percentage difference in the modelled spectrum when Fe K$\alpha$ lines are included with conservative assumptions (see text), compared to the best model without the emission lines. The peak (and bulk of this feature) is due to type 2 sources at a characteristic redshift of 0.7, as described. The much weaker second peak corresponds to the break at $z$=1.5 of type 1 sources. The sharpness of the peaks is partly due to the fact that a low-resolution grid was used for modelling the lines.
} \label{fig:residuals} \end{center}
\end{figure}

The lines are assumed to be narrow gaussian profiles with a fixed width $\sigma$=10 eV (In practice, the resolution of the energy grid was such that the lines had a relatively sharp peak). The rest-frame equivalent-widths (EW) of lines in type 1 sources has been taken to be 150 eV \citep{georgefabian91}. For type 2 sources with \lognh$>$23.25, we match to the relation found by \citet{leighycreighton} where the EW is 180 and 900 eV for \lognh=23.75 and 24.25 respectively, while for sources with \lognh=22.25...23.25, they found an EW less than 150 eV. Thus, for these sources, we fix EW=150 eV as well. For the highest column densities, \citeauthor{leighycreighton} found column-densities greater than 10 keV. To be conservative, we choose to fix the Fe K$\alpha$ EW to 6 keV for \lognh=24.75 (this was the maximum EW found in a sample of Seyfert 2s by \citealt{levenson02}). This is the strength of the line generated from fluorescence in the obscuring matter itself. Fig~\ref{fig:residuals} shows the effect produced. Compared to the best model without the Fe line included, we find a maximum difference of more than 3 per cent close to 6.4$/(1+z)$=3.8 keV, where $z$=0.7 is the characteristic source redshift. Though inclusion of cluster emission may dilute the EW of this feature on the XRB spectrum itself, especially at softer energies, we emphasize that such a weak feature should be searched for, especially if the (observed) redshift distribution peaking at $z$=0.7 is to be believed.



\subsection{Obscured AGN contribution to IR counts}

Finally, we make use of the correlation between X-rays and the IR to make predictions of type 2 AGN source counts at 70 \micron\ (the MIPS instrument aboard the forthcoming \sirtf\ mission has a 70 \micron\ 5$\sigma$ sensitivity of 1.3 mJy in 500 s; e.g., \citealt{brandl}). 

For $k$-corrections, we construct models with the public radiative transfer code DUSTY \citep{dusty}, with all the assumptions of \citet{wfg}. The main variable values assumed are : $r_{\rm out}/r_{\rm in}$=500, the relative scales of the spherical dust distribution; $T_{\rm in}$=1000 K, the dust temperature at $r_{\rm in}$ and $\tau$(0.3 \micron)=10...10000, the optical-depth. These then correspond to type 2 X-ray sources with \lognh$\approx$22...25, assuming a Galactic dust:gas ratio and solar abundances. DUSTY cannot currently compute a radiative transfer solution through a torroidal dust geometry. Although we are thus biased towards fainter sources (and our estimates will be lower-limits), we note that new observations are finding many objects which have a higher covering fraction than that inferred from a torus. Absence (or weakness) of optical and near-infrared emission lines in many newly-discovered \c\ and \xmm\ sources suggests a dust covering fraction large enough to completely obscure each source. Indeed, the flat shape of the observed XRB spectrum suggests a very large covering fraction (e.g., \citealt{fi} deduce this to be approximately 85 per cent). Moreover, good fits to the broad-band SEDs of optically-red and X-ray hard sources have been obtained under the assumption of complete obscuration and / or very low scattering fraction \citep{wfg, franceschini02b}. The case of spherical obscuration may be valid for a non-negligible fraction of distant type 2 AGN. We return to this point in the next section.

\begin{figure}
  \begin{center}
\includegraphics[angle=90,width=8.5cm]{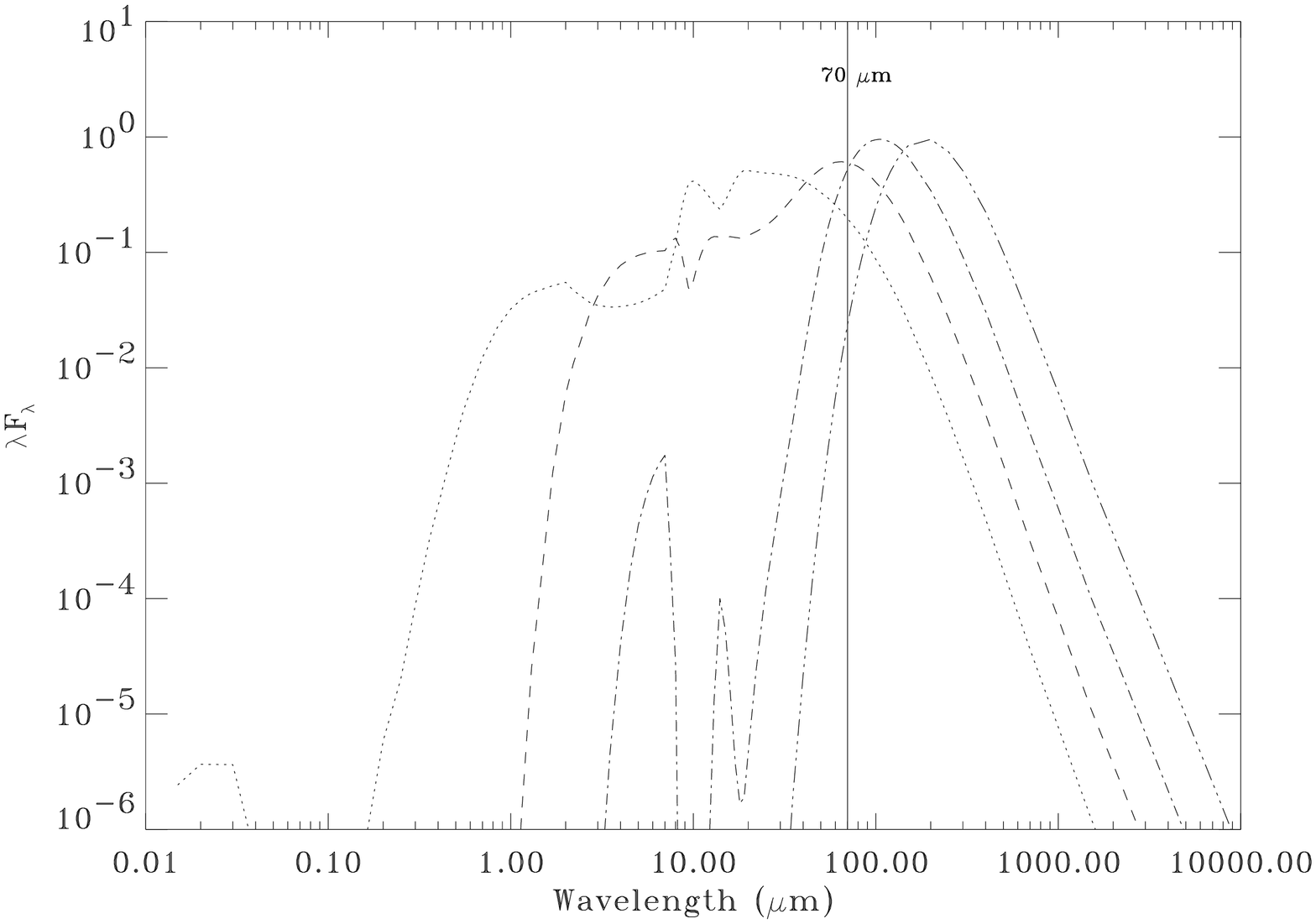}
\includegraphics[angle=90,width=8.5cm]{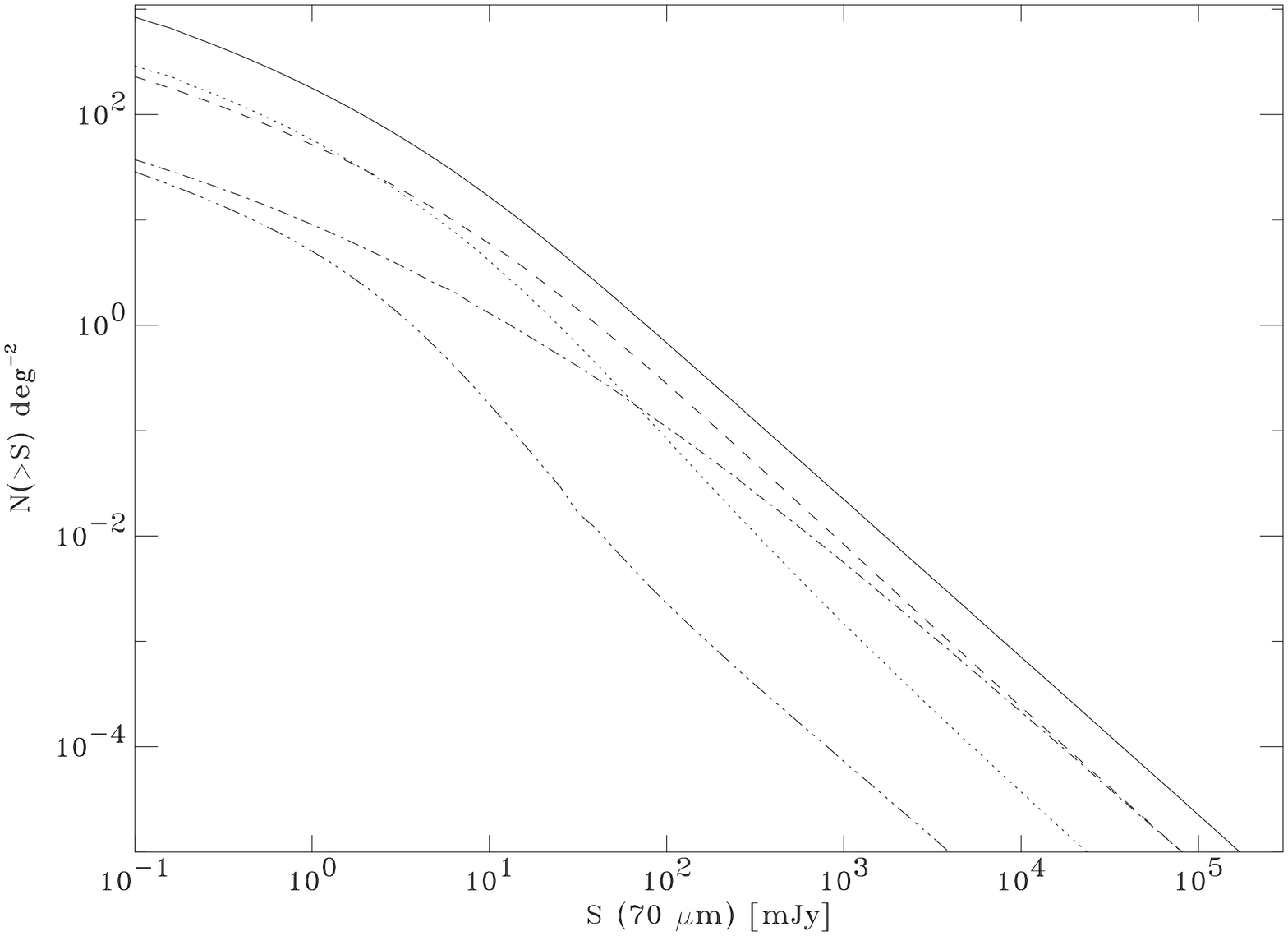}
  \caption{(Top) DUSTY spectra with the parameters described in the text and X-ray \lognh=22 (dotted), 23 (dashed), 24 (dot-dashed), 25 (dot-dot-dot-dashed), corresponding approximately to $\tau_{\rm 0.3 \mu m}$=10...10000. (Bottom) Type 2 AGN source contribution to the 70 \micron\ source counts with the $k$-corrections of the spectra in the top figure, and density distribution of the model described in \S~\ref{section:results1}. Total counts -- solid line; individual contributions -- as in the top figure.} \label{fig:sirtf} \end{center}
\end{figure}

To determine the IR:X-ray ratio, we determine the ratio of the bolometric to 70 \micron\ (monochromatic) luminosity for each SED, and convert this to an intrinsic 2--10 keV luminosity using the ratio found by \citet{elvis94} for a sample of local, unobscured, radio-quiet quasars. We can now use the type 2 X-ray LLF and the evolution derived for our illustrative model and predict 70 \micron\ source counts above a limiting flux of 0.1 mJy. These are shown in Fig~\ref{fig:sirtf}. Comparing this with the prediction made by \citet[][ their Fig 20]{franceschini01_long}, we find that at the faintest fluxes, type 2 AGN will constitute less than 1 per cent of the cumulative IR counts. However, our estimate must be treated as a lower limit since we have not modelled starburst contributions to the SED (which would make the $k$-correction higher 
-- especially the most obscured sources -- and increase number counts significantly).

If, on the other hand, the dust:gas ratio is lower than the Galactic value \citep[as has been seen at low redshift; e.g., ][]{maiolino01}, the IR counts will increase significantly. For instance, assuming for simplicity that the typical extinction due to dust is uniformly lower by a factor of 10 \citep[e.g., ][]{granato97} than that expected from the observed amount of X-ray obscuration (and if the dust:gas ratio is Galactic), we obtain number counts which are higher by a factor of two at the faintest flux level.

\section{Discussion}

We have presented a model in which a distinction is made between type 1 and type 2 X-ray source distribution and evolution, and type 2 sources follow a (rescaled) luminosity function of MIR sources. With simple power-law emissivity evolution, we have obtained relatively good fits to the XRB spectrum over 5$\sim$100 keV. 
Some density evolution is required to match the increasing integral source counts at faint fluxes. We have not fit parameters relevant to type 1 sources, although their relative fraction does vary with changing $\beta$. 3--5 per cent of IR sources are found to be obscured AGN locally (4 per cent for the model in Fig~\ref{fig:xrbspec}) and this increases to 7--13 per cent (of the local IR number density) at $z=0.7$, depending on the value of $\beta$. We note that the illustrated model is not the \lq best\rq\ one in the sense of having the minimum deviations from the XRB spectrum of all possible models. However, this does have moderate success in accounting for all observational constraints (the XRB spectrum, X-ray number counts and observed peak of the redshift distribution). The main uncertainties remain the exact LF of type 2 sources and the IR:X-ray luminosity ratio for each \nh\ sub-class.

We find residuals on the XRB spectrum due to intrinsic Fe emission which are not significantly smeared out by the peaked redshift distribution. This feature is weak and at the level of a few per cent. It lies close to the lower range (3 keV) of energies over which \citet{marshall80} found no significant residuals; its non-detection, therefore, is not necessarily in conflict with our model, and further effort should be made to either confirm or rule out this feature observationally.

We have extended the work of F02 by using a distribution of absorbed X-ray spectra with full Thomson scattering cross-sections. In so doing, however, we have less certainty on the IR:X-ray luminosity ratio for each $N_{\rm H}$ sub-class. This could explain our lower estimate (4 per cent) of the local fraction of IR sources which are obscured AGN. We have investigated a range of parameter space for the type 2 evolutionary parameters, and commented on the need for various classes of sources. Differences with F02 in the determined redshift distributions are partly due the different IR LFs used as a starting point and to differences in cosmology. For instance, the number of type 2 sources per deg$^2$ in a non-zero lambda cosmology with $\Omega_{\rm M}=0.3$ and $\Omega_{\Lambda}=0.7$ is lower by about 15 per cent at $z=4$ (refer to Fig~\ref{fig:nz}; this effect is larger at high redshifts due to the rapidly decreasing comoving volume). Additionally, adopting a higher flux threshold of $8\times 10^{-16}$ [cgs] leads to 40 per cent less counts at $z=4$ (the limiting fluxes of the {\sl observed} redshift distribution vary between the samples, which makes exact comparison difficult). 


The intimate connection between star-formation and AGN accretion activity has been known for some time now, especially in the most massive galactic systems (see e.g., \citealt{genzel98} and \citealt{magorrian98} for different lines of evidence). Phenomenological models which self-consistently account for the symbiotic growth of a black hole and its host bulge have been explored (e.g., \citealt{f99}; \citealt{silkrees}) and can reproduce observed correlations such as the mass-dispersion relation of \citet{gebhardt00} and \citet{merrittferrarese01}. \citet{haiman00} have shown that, in a primordial cloud, enhanced production of photoelectrons due to emission of hard X-rays from a miniquasar facilitates the formation of $H^-$, leading to increased abundance of $H_2$ which allows cooling and collapse of dense virialized clouds, possibly leading to star-formation. This effect may be stronger for halos with higher metallicity.

What the new deep field observations suggest, is a universal peak phase of accretion activity at $z\sim 0.7$, which is delayed from the phase of unobscured quasar activity peaking at or beyond $z\sim 1-2$. Recent determination of the star-formation history also shows a flattening beyond $z\sim0.7$ (\citeauthor{ce01}), and it has already been suggested that an increased rate of mergers at these intermediate redshifts could supply additional gas to trigger as well as feed nuclear star-formation and also lead to obscured accretion. While an increase in the merger rate has been observed to $z$=1 \citep{lefevre00}, a detailed study of this merger rate would be essential to accurately determine the higher redshift connection with feeding obscured AGN. We also note that \citet{xu00} found a higher \zcut=1.5, based on the direct (rather than derived) 15 \micron\ \iso\ LF. Follow-up observations of distant \sirtf\ sources may resolve this issue within the next few years.



The local mass density in black holes based on the redshift distribution of our model is found to be $\rho_{\rm BH}=7.6\times 10^{5} (\eta/0.1)^{-1} (f/0.03)^{-1}$ \Msun\ Mpc$^{-3}$, predicting a bolometric accretion energy density of $1.3\times 10^{58} (f/0.03)^{-1}$ erg Mpc$^{-3}$ ($\eta$ is the accretion efficiency and $f$ is the 2--10 keV : bolometric correction factor adopted from the radio-quiet quasar compilation of \citealt{elvis94}). 

This is more than a factor of 2 larger than the mass density found by \citet{fi} from simple absorption correction of the observed XRB spectrum (adopting a median redshift of 0.7), and less by the same factor as compared to the upper range of densities quoted by \citet{elvis02}, based on better estimates of the bolometric correction, $<z>$=2 and assuming $\eta$=0.1 (We compare with the upper limit since our estimate would also increase with the new corrections). \citet{yu_tremaine02} and \citet{merritt_bhdem} find densities which are lower by a factor of 2--4 (after homogenizing the adopted cosmology) based on the locally observed $M_{\rm BH}-\sigma$ relation. \citeauthor{yu_tremaine02} conclude that efficient accretion in {\sl optically-bright} QSOs can account for the calculated $\rho_{\rm BH}$. In this regard, we note that a significant fraction of our mass density is due to obscured accretion at high redshifts (we integrate up to \zmax=4.5), and if we adopt a \zmax=1.5 for type 2 sources, $\rho_{\rm BH}$ drops to $4.9\times 10^{5}$ \Msun\ Mpc$^{-3}$ without affecting the XRB spectrum by the same factor. However, the flat slope of the XRB spectrum unambiguously suggests the presence of powerful objects which are highly obscured, and this is difficult to reconcile with the spectra of type 1 QSOs. 
Again, the results from \sirtf\ surveys \citep[e.g., ][]{lombardi01, rieke01} should provide important constraints.

\citet{f98} discussed the possibility that nuclear star-formation could provide the energy required to keep a circum-nuclear torus puffed up with the large covering fraction implied by the XRB spectrum. However, the complete absence of emission lines (from either star-formation or AGN activity) in both the optical (e.g., \citealt{hasinger02}) and the near-infrared \citep{g02} in many X-ray obscured sources suggests that star-formation is either absent, or highly reddened. Although larger samples of \c\ and \xmm\ follow-up spectra are required for quantitative analysis, it seems clear that obscuration at high redshift must be orientation-independent and that the simple AGN unification based on orientation cannot hold at increasing redshift. Whether the difference between distant type 1 and type 2 AGN is due to nature (the conditions during formation) or nurture (effects of the environment, e.g., mergers) or a mixture of these effects, needs to be understood.

\section{Acknowledgments}

We thank Richard Wilman for making the monte carlo code available to us. PG thanks the Sir Isaac Newton Trust and the Cambridge Overseas Research Trust for financial assistance. ACF acknowledges the Royal Society for support. We also thank the referee for useful suggestions.

\bibliographystyle{mnras}                       
\bibliography{mnrasmnemonic,MC815}

\end{document}